\documentclass[preprint,showpacs,preprintnumbers,amsmath,amssymb,11pt]{revtex4}
\usepackage{graphicx}
\usepackage{dcolumn}
\usepackage{bm}
\usepackage{color}
\newcommand{\be}{\begin{equation}}
\newcommand{\en}{\end{equation}}
\newcommand{\bea}{\begin{eqnarray}}
\newcommand{\ena}{\end{eqnarray}}
\begin{document}
\title{Intermediate inflation under the scrutiny of recent data}
\author{Sergio del Campo}
\email{sdelcamp@ucv.cl}
\affiliation{ Instituto de F\'{\i}sica, Pontificia Universidad Cat\'{o}lica de Valpara\'{\i}so, Casilla 4059, Valpara\'{\i}so, Chile.}
\date{\today}
\begin{abstract}
We use the flow equations to determine the different hierarchy
Hubble parameters as a function of the number of e-folds for
intermediate models in single-field inflation. The obtained
expressions allow us to determine at second order in the hierarchy
Hubble parameters different observational parameters. We
distinguish the scalar spectral index, its running and the
tensor-to-scalar ratio, among others. Recently, it has been
noticed that measurements released by Planck, combined with the
WMAP large-angle polarization are in tension with this sort of
model. Here, we show in detail why this occur. The conclusions do
not change even when the recent BICEP2 data are included.

\end{abstract}
\maketitle

\section{Introduction}


In single-field inflationary universe theories there exist three
ways to address their study. One of them is the usual slow-roll
approach which has been considered by many researcher from its
beginning\cite{G81,AS82,L82,L83,ST84,L84,SB90,LL92,LPB94}.
This approximation  puts generic restrictions on the effective
inflaton scalar potential\cite{AFG91}, but it has been quite successful
in describing the main characteristics of inflationary scenarios. However,
it has been shown that the
slow-roll approximation is invalid for all models in which the
scalar field interacts with itself\cite{MUW85,M85}, specifically
when the potential presents a inflexion point, where a violation of
the slow roll approximation is encountered. In this
approach the expansion of the universe is governed by the
effective scalar field potential, $V(\phi)$ ($\phi$ here is the {\it scalar inflaton
field}), where the kinetics term is much smaller than the potential
energy. Added to this approximation it is also assumed that
$\mid \frac{d \ln{\dot{\phi}}}{dt}\mid\, \ll\, H$, which
simplifies the Klein-Gordon equation to $3\,H\,\dot{\phi}\,\approx
\frac{d V(\phi)}{d\phi}$, where dots represent derivatives with respect
to the cosmological time $t$. To make this scheme to work it is needed
to provide an explicit expression for the effective inflaton
potential $V(\phi)$. The picture here it is that during inflation the
inflaton field slowly rolls down to the minimum of the scalar
potential.

Essentially, we can observe that the slow-roll inflationary
scenario is intimated related to the inflaton scalar field
potential, $V(\phi)$. Since, most of the observational quantities
(such that the spectral scalar index, $n_s$, the running scalar
index, $\alpha_{s} \equiv \frac{d n_s}{d \ln k}$, the
tensor-to-scalar ratio, $r$, the spectral tensor index, $n_T$, etc.)
are expressed in terms of the
scalar potential and its derivatives, via the different slow-roll
parameters. Therefore, we expect that the observational data will
put strong constraint on the shape of the potential. In fact, one
of the predictions of the slow-roll inflation is that the form of
the potential should be extremely flat and smooth in the case of
minimally coupled scalar field. Due to this, it is possible that
the slow-roll approximation may breaks down for some range in the
values of the inflaton scalar field where the scalar potential does
not present these qualities. Moreover, it has been noticed that in
hybrid inflation the slow-roll approximation breaks down at all
points in the evolution of the scalar field\cite{G-BW96,K97}. More
dramatic situations can occur. For instance, in the model discussed
by Starobinsky\cite{S92}, characterized by a scalar potential that
presents a sudden change in its slope at $\phi=\phi_0$, where the inflaton potential
becomes $\displaystyle
V(\phi)=\left\{
           \begin{array}{rl}
           V_0 + A_+(\phi - \phi_0) &\mbox{ for $\phi > \phi_0$} \\
            V_0 + A_-(\phi - \phi_0) &\mbox{ for $\phi < \phi_0$}
             \end{array} \right..$
The change in the slope of this potential may be sufficiently
abrupt so that the slow-roll approximation can be violated, and for
$A_+ > A_- > 0$ the field enters in a  "fast-roll" solution,
characterized by the expression $ \ddot{\phi} = -3 H
\dot{\phi}$\cite{LL01,LSWL01}.

Intermediate and power law inflations provide exact solutions for
the time evolution of cosmological perturbations, and inflation can
occur although the slow-roll conditions are violated. It is
therefore interesting to investigate the exact predictions out of
the slow-roll predictions. For instance, in \cite{SM98} was found
that the consistency relation obtained from the slow-roll
approximation, $C^T_2 /C^S_2 \thickapprox -6.93n_T$, may differ
considerably from the exact result. In power law inflation, where
the scale factor evolves as $a \backsim t^p$, with  $p > 1$, the
exact and the slow-roll results differ by a factor $F(n_T)/(1 -
n_T/2)$, where $F(n_T)$ denotes a numerical integration. For
instance, if $p$ is taken to be two the error is found to be as
high as a $34\%$.

Another way to study inflationary universe models out of the
slow-roll approximation is given by the Hamilton-Jacobi
approach\cite{GS88,Letal97,dC12,dC13,dC14}. In this schema the basic
quantity results to be the Hubble parameter which is given in
terms of the inflaton field, $H(\phi)$, the so called generating
fuction\cite{L91}. In this approach the form of the potential is
deduced, and since we are out of the slow-roll approximation,
application to the final period of inflation is possible. In this
period the kinetic term associated to the inflaton field in the
Friedmann equation becomes important when compared with the scalar
potential. In this way, this approach becomes useful  when
studying the final stage of inflation in which the reheating phase
occurs. In the same way, this approach can be applicable to models
that cannot be studied within the slow-roll approximation, for
instance in models where the problem of a large slow-roll
parameter $\eta$ of some supersymmetry-inspired inflationary
models arises,  and it can be applied to small scales that leave
the horizon at times close to the end of inflation, where, as we
have mentioned,  the slow-roll approximation necessarily breaks
down. In this approach we can take the inflaton scalar field,
$\phi$, as a {\it time variable}, and for that, we claim that this
field increases monotonically, i.e. its time derivative,
$\dot{\phi}$, should not change of sign along the inflationary
phase.

The third way of studying inflationary universe models is through
the introduction of certain parameters which are subtended by a
sort of order imposed on them\cite{HT01,K02}. Each of these
hierarchy parameters is characterized by its dependence on the
order of the scalar field derivative of the Hubble ratio,
$H(\phi)$. Prior to the introduction of the approach as such let
us bring out the so-called {\it first Hubble hierarchy parameter}
defined by \be \epsilon_{_{H}} \equiv - \frac{d \ln {H}}{d \ln
{a}} = \left(\frac{m_{_{Pl}}^2}{4
\,\pi}\right)\,\left(\frac{1}{H}\,\frac{d\,H}{d\,\phi}\right)^2,\label{e} \en
where $m_{Pl}$ represents the Planck mass.

The previous fundamental quantity becomes defined from the the acceleration
equation for the scale factor
\be \frac{\ddot{a}}{a}=
H^2\left[1-\epsilon_{_{H}}\right].\label{dda} \en
We observe that during inflation, i.e., when $\ddot{a}$ is
positive, this parameter satisfies the bound $\epsilon_{_{H}} \leq
1$, where the equality is obtained at the end of the inflationary
period.

As we did with the previous definition we can introduce similar
parameters which are also called {\em Hubble hierarchy parameters}.
The {\em second Hubble hierarchy parameter}, $\eta_{_H}$, is
defined by
\be \eta_{_H} \equiv -\left(\frac{d\,\ln
\left(\frac{dH}{d\phi}\right)}{d\,\ln a}\right)=
\frac{m_{Pl}^2}{4\,\pi}\,\left[\frac{1}{H}\,\frac{d^2H}{d
\phi^2}\right]. \label{eta} \en
Similarly, we can introduce the {\em third Hubble hierarchy
parameter}, $\xi_{_{H}}^2$, defined by
\be \xi_{_{H}}^2 \equiv
\left(\frac{d\,\ln\left(\frac{dH}{d\phi}\right)}{d\,\ln
a}\right)\left(\frac{d\,\ln
\left(\frac{d^2H}{d\phi^2}\right)}{d\,\ln a}\right)=
\left(\frac{m_{_{Pl}}^2}{4\,\pi}\right)^2\,
\left[\frac{1}{H^2}\,\frac{d H}{d \phi}\,\frac{d^3H}{d
\phi^3}\right].\label{xi} \en
We can extend these definitions to higher derivatives of the
Hubble parameter, so that we can define in general
\begin{eqnarray} ^l\lambda_{_H}& \equiv & (-)^l\left(\frac{d
\ln\left(\frac{dH}{d\phi}\right)}{d \ln a}\right) \left(\frac{d
\ln\left(\frac{d^2H}{d\phi^2}\right)}{d \ln a}\right)\cdots  \cdots \cdots
\left(\frac{d \ln\left(\frac{d^lH}{d\phi^l}\right)}{d \ln a}\right) \nonumber \\
&= &\left(\frac{m_{_{Pl}}^2}{4\,\pi}\right)^l\left[
\frac{1}{H^{^{l}}}\,{\left(\frac{d H}{d
\phi}\right)^{l-1}}\,\frac{{d\,}^{^{l+1}}H}{d
\phi^{^{l+1}}}\right],\hspace{2.8cm} (l\geq 1),\label{lambda}\end{eqnarray}
where we see that for $l=1$ it gives $^1\lambda_{_H} \equiv \eta_{_{H}}$ and for $l=2$
it provides $^2\lambda_{_H} \equiv \xi_{_{H}}^2$, etc.

In this approach the set of equations  is based on
derivatives of the different Hubble hierarchy parameters $^l\lambda_{_H}$ with
respect to the {\it e-folds number}, $N$. Here, the parameter
$N$ is interpreted as the number of e-folds before inflation ends.
This quantity is defined as $ N(t) \equiv \ln
{\frac{a\left(t_{end}\right)}{a(t)}},\label{NCh} $
where $a\left(t_{end}\right)$ represents the scale factor
evaluated at the end of inflation. Usually this quantity is written as
$ N = - \int_t^{t_{end}} {H \,dt}$.

In this paper we would like to study intermediate inflationary universe
models by taking into account the hierarchy flow equations. After giving a brief
introduction to intermediate inflation, we solve the set of hierarchy equations
in such a way that, as we will see, it will be necessary
to  determine the {\it first Hubble hierarchy parameter}, $\epsilon_{_{H}}$, since the other
high parameters will be determined from this one. After doing this, we will determine different
observational quantities in terms of these parameters. We will obtain these quantities at
second order of the Hubble hierarchy parameters. Then,
we will contrast these parameters
with the corresponding observational data released by Planck, together with
the WMAP large-angle polarization observations\cite{P13}. Interesting results the data expressed
in the $n_s/r$ plane, where the different models make predictions on it. Certain increasing
precision will help in selecting the appropriated model.
On the other hand, very recently, the ground based BICEP2 experiment\cite{B14}
announced the detection for the first time of primordial B-mode polarization
on the cosmic microwave background radiation. This significant measurement implies
that the tensor-to-scalar ratio, $r$, present a non-zero value at seven sigma,
whose value (combination of Planck + WMAP9 + HighL + BICEP2) results to be $r = 0.2^{+0.07}_{-0.05}$.
Not only this discovery should be confirmed by another ground based or satellite experiment,
but also this new experiment should determine the amplitude of
the corresponding gravitational wave signal. Perhaps, Planck satellite will help on this task.
As a result of this finding the $n_s/r$ plane becomes modified. In this article we
would like to use the current available cosmological data to look at the feasibility
of the inflationary model called intermediate.

\section{A brief introduction to intermediate inflation}

Intermediate inflationary universe models was introduced as an
exact solution for a particular scalar field potential of the type
$ V(\phi) \propto \phi^{-4(f^{-1}-1)}$ \cite{var05},
where $f$ is a free parameter which ranges as $0 < f < 1$. With
this sort of potential it is possible in the
slow-roll approximation to have a spectrum of density perturbations
which presents an exact scale-invariant spectral index, i.e. $n_s = 1$, the
so-called Harrison-Zel'dovich spectrum of density perturbations.

The main motivation to study intermediate inflationary model
becomes from string/M theory. This theory suggests that in order
to have a ghost-free action high order curvature invariant
corrections to the Einstein-Hilbert action must be proportional to
the Gauss-Bonnet (GB) term \cite{BS85}. GB terms arise naturally
as the leading order of the expansion to the low-energy string
effective action, where is the inverse string tension\cite{KM07}.
This kind of theory has been applied to possible resolution of the
initial singularity problem\cite{ART94}, to the study of
Black-Hole solutions\cite{var06}, accelerated cosmological
solutions\cite{var07}, among others. In particular, it has been
found that for a dark energy model the GB interaction in four
dimensions with a dynamical dilatonic scalar field coupling leads
to a solution of the form $a(t) = a_0
\exp{\left[\left(\frac{2}{\kappa\,n}\right)\,t^{\frac{1}{2}}\right]}$
\cite{S07}. Here, $\kappa = 8\,\pi\,G$ and $n$ is an arbitrary
constant. Actually, this kind of behavior of the scale factor is
what characterizes the  intermediate inflation, in which the scale
factor evolves as \be a(t) =
a_0\exp{\left(At^f\right)},\label{a}\en where $A$ is a positive
constant and $f$ was introduced above. In this way, the expansion
of the universe is slower than standard de-Sitter inflation ($a(t)
= \exp{(Ht)}$), but faster than power law inflation ($a(t) = t^p;
p > 1$). Thus, the idea that inflation, or specifically,
intermediate inflation, comes from an effective theory at low
dimension of a more fundamental string theory is in itself very
appealing. Thus, in brane universe models the effective theories
that emerge from string/M theory lead to a Friedmann equation
which is proportional to the square energy density, on the one
hand, and an evolving intermediate scale factor, in addition, it
makes interesting to study their mixture by itself, i.e., an
intermediate inflationary universe model in a brane world
effective theory. In it was shown that the combination $n_s = 1$
and $r > 0$ is given by a version of the intermediate inflation
model in which the scale factor varies as $a\sim
\exp(\frac{t}{t_o})^{2/3}$ in which the slow-roll approximation
was used\cite{BLP06}. In the same reference the status of this
model was evaluated in general terms in light of the WMAP3 data.
But, as was described in the introduction, new data are accessible
nowadays, and we pretend to use them in order to appraise
intermediate inflation.

In the following we will describe the main results that occur in
intermediate inflation in the most simpler model in which a single
scalar inflaton field is present.

Our basic equations are the Friedmann equation which we take to be
\be H^2 = \left(\frac{8 \pi}{3
m_{Pl}^2}\right)\left[\frac{1}{2}\dot{\phi}^2 + V(\phi)\right],
\label{h} \en
 together with the energy density conservation equation
\be \ddot{\phi} + 3 H \dot{\phi} = -\frac{d\,V(\phi)}{d\,\phi}. \label{phi}
\en
It is possible to find exact inflationary universe solutions
from this set of equations
by giving the functional form of the Hubble
parameter in term of the inflaton field, i.e. $H(\phi)$ by making use
of the Hamilton-Jacobi approach.
From now on, we will adopt this type of approach.

From expressions (\ref{h}) and (\ref{phi}) yield to
\be \dot{\phi}^2 = -\,\left(\frac{m_{Pl}^2}{4 \pi}\right)\,\frac{d\,H}{dt}. \label{dphi}\en
Now, equation (\ref{a}) allows to obtain the Hubble factor as a
function of the cosmological time, and thus the latter
equation can be integrated for obtaining the scalar field as a function of time.
The result is given by
\be
\phi(t)=\left(\frac{m_{Pl}^2}{4\,\pi}\,\beta\,A\,t^f\right)^{\frac{1}{2}},
\label{phit} \en
where $\beta \equiv 4\left(f^{-1}-1\right)$ is a dimensionless positive constant.
This latter expression can
be inverted to obtain the cosmological time as a function of the scalar field
which can be used for obtaining the generating function. We get that in
this case the generating function becomes
\be
H(\phi)=\left(\frac{m_{Pl}^2}{4\,\pi}\,\beta\,A\right)^{\frac{\beta}{4}}A\,f\,\phi^{-\frac{\beta}{2}}.
\label{Hphi} \en
From this latter expression we can get the scalar inflaton potential,
which becomes expressed  by
\be V(\phi) = \frac{3 m_{Pl}^2}{8
\pi}\,H^2\left[1-\frac{m_{Pl}^2}{12 \pi}
\left(\frac{1}{H}\,\frac{d\,H}{d\,\phi}\right)^2\right]=
\mathcal{V}_0\,\phi^{-\beta}\left[1-\left(\frac{m_{Pl}^2}{48\,\pi}\right)
\left(\frac{\beta}{\phi}\right)^2\,\right],\label{5}\en
where the constant $\mathcal{V}_0$ becomes given by $\displaystyle \mathcal{V}_0
\equiv
\frac{3}{2}\left(\frac{m_{Pl}^2}{4\,\pi}\right)^{\frac{\beta}{2}+1}f^2\,\beta^{\frac{\beta}{2}}\,A^{\frac{2}{f}}$.

At this point some comments are in order. First, the potential
presents extreme values at
$\phi_{_{\pm}}=\pm\sqrt{\left(1+\frac{2}{\beta}\right)B}$, with
$B=\left(\frac{m_{Pl}^2}{48\,\pi}\right){\beta}^2$ and in the following
we will analyze the scalar potential in the ranges $\phi > 0$ and $\phi < 0$,
separately. Second, it
is found that the second derivative of the potential evaluated at
the previous extreme points becomes $\displaystyle \left.\frac{d^2
V(\phi}{d\phi^2}\right|_{\phi=\phi_{_{\pm}}}=-\frac{2\beta^2}{(2+\beta)B}\left[\pm
\left(1+\frac{2}{\beta}\right)B\right]^{-\frac{\beta}{2}}$. We see that
$\phi=\phi_{_+}$ represents an maximum extreme point of the scalar potential
for any value of the $\beta$ parameter. Therefore, no minimum it is found
in the range in which the inflaton field is positive. The situation in which the
point $\phi=\phi_{_{-}}$ is concerned, is more subtle.  Note here that it is possible
to have an extreme value (maximum or minimum) depending of the value of $\beta/2$.
Firstly, if this value is integer odd, then we find that the second
derivative of the potential becomes positive and that the potential present a minimum
at the point $\phi=\phi_{_{-}}$. But, if the parameter  $\beta/2$ is integer even again
it is found a maximum point in the scalar potential. The situation in which this
parameter becomes non-integer can not be since the second derivative of
the potential becomes an imaginary number.

In light from the previous analysis we see that the negative
range for the inflaton field
becomes more restricted than the case in which
the range becomes positive. In our analysis
we would like to relax the $\beta$ (or $f$) parameter, so that,
in the following we take the inflaton field within the positive range.
The price that we pay here is that, since the scalar potential
does not present a minimum, then it is not possible  to finish
the inflationary period\cite{minimum}.

\section{The hierarchy flow equations}

The corresponding hierarchy flow equations become expressed as\cite{K02}
\begin{eqnarray}\label{flo} \frac{d\epsilon_{_{H}}}{d N} & = & \epsilon_{_{H}}(\sigma + 2
\epsilon_{_{H}}), \nonumber \\
\frac{d \sigma}{d N} & = & -5 \epsilon_{_{H}}\sigma -12
\epsilon_{_{H}}^2 + 2\xi_{_{H}}^2 , \\
\frac{d\, {^l\lambda}_{_{H}}}{d N} & = & \left[\frac{l-1}{2}
\sigma + (l-2)\epsilon_{_{H}}\right]\,{^l\lambda_{_{H}}} +
{^{l+1}\lambda_{_{H}}},\hspace{1cm} (l\geq 2)\nonumber
  \end{eqnarray}
where $\sigma$ is defined as $\sigma = 2\,\eta_{_{H}}-4\,\epsilon_{_{H}}$.

As specified in the introduction is only necessary to know the first
hierarchy parameter, $\epsilon_{_{H}}$, as a function of the number of e-folds, $N$,
and then from it, the other parameters are determined by using the previous set of equations.
So, let us assume for a moment that we know the parameter $\epsilon_{_{H}}$
as a function of $N$. Then, from
the first equation of the set (\ref{flo}) we get that
\be \eta_{_H} =  \epsilon_{_{H}}+\frac{1}{2}\left(\ln
\epsilon_{_{H}}\right) ', \label{etaepsilon} \en
where the prime represents a derivative with respect to the number of e-folds, $N$.

Similarly, from the second equation of the set  (\ref{flo}) we obtain
\be \xi_{_{H}}^2
=\epsilon_{_{H}}^2+\frac{3}{2}\epsilon_{_{H}} '
+\frac{1}{2}\left(\ln \epsilon_{_{H}}\right)''. \label{xiepsilon} \en

Analogously, we get for $l=3$ the following expression
\be
{^3\lambda}_{_{H}}=\epsilon_{_{H}}^3+3\epsilon_{_{H}}\,\epsilon_{_{H}}'
-\frac{3}{4}\frac{1}{\epsilon_{_{H}}}\left(\epsilon_{_{H}}'\right)^2+\frac{3}{2}\epsilon_{_{H}}''
+\frac{1}{2}\left(\epsilon_{_{H}}-\frac{1}{2}\left(\ln\epsilon_{_{H}}\right)'\right)\left(\ln\epsilon_{_{H}}\right)''+\frac{1}{2}\left(\ln
\epsilon_{_{H}}\right)''', \label{3Lepsilon} \en
and similarly the other high parameters can be obtained. As we see, we can say that any of the high
hierarchy parameters can be determined from the first Hubble hierarchy parameter,
just using the set of flow equations.

\section{The hierarchy Hubble parameters for intermediate inflation}

\subsection{The generating function in terms of N}

The generating function as a function of the
number of e-folds we take to be given by
\be H(N)=H_e\left(1-\frac{4\,N}{\beta}\right)^{-\frac{\beta}{4}},
\label{HN} \en
where $H_e$ represents the value of the Hubble parameter at the
end of inflation (when $N=0$) and is given by $\displaystyle H_e =
(Af)
\left[\frac{4A}{\beta}\left(\frac{m_{Pl}^2}{4\pi}\right)^2\right]^{\frac{\beta}{4}}$.
Since $H$ must be a positive real quantity we need to impose the constraint $\beta > 4N$, which for
$N=(50,60,70)$, the usual values of the required  number of e-folds, implies that
$\beta$ should be grater than $(200,240,280)$.
This is equivalent to taking the inequality $f<(0.019,0.016,0.014)$
for values of the parameter
$f$.

Since we know how the scale factor evolves with the cosmological time,
we get the Hubble parameter as a function of time.
By using equation (\ref{HN}) we obtain the time as a function of $N$
\be
t(N)=\left[\frac{\beta}{4A}\left(1-\frac{4N}{\beta}\right)\right]^{\frac{1}{f}}.
\label{tN} \en
Note that the period of inflation becomes given by $t_{e}\equiv
t(N=0)=\left(\frac{\beta}{4A}\right)^{\frac{1}{f}}$. On the other hand, expression (\ref{tN}) yields to
\be
\phi(N)=\frac{\beta}{2}\sqrt{\left(\frac{m_{Pl}^2}{4\pi}\right)\left(1-\frac{4N}{\beta}\right)},
\label{phiN} \en
where $\phi_e \equiv\phi(N=0) = \frac{\beta}{2}\sqrt{\frac{m_{Pl}^2}{4\pi}}$
is the value of the inflaton field
at the end of inflation. It is not difficult to see that when equation
(\ref{phiN}) is substituted into equation (\ref{HN})
we obtain the generating function (\ref{Hphi}).

\subsection{The hierarchy Hubble parameters in terms of N}

We can use expression (\ref{HN}) for getting the first hierarchy
parameter, $\epsilon_{_{H}}$, since it is given by $\displaystyle
\epsilon_{_{H}} = \frac{d}{dN}\left(\ln H(N)\right)$. This becomes
\be
\epsilon_{_{H}}(N) = \left(\frac{1}{1-\frac{4N}{\beta}}\right).
\label{epsilonN}
\en
Note that at the end of inflation, i.e. when $N=0$,
this parameter takes the value one, as it should be.

The other parameters, $\eta_{_{H}}$ and $\xi_{_{H}}^2$, can be obtained by using expressions
(\ref{etaepsilon}) and (\ref{xiepsilon}), respectively. They become
\bea \eta_{_{H}}(N)=
\left(\frac{1+\frac{2}{\beta}}{1-\frac{4N}{\beta}}\right), & \mbox{and} & \hspace{0.7 cm}
\xi_{_H}^2(N)=\frac{1+\frac{6}{\beta}+\frac{8}{\beta^2}}{\left(1-\frac{4N}{\beta}\right)^2},\label{etaN}
\ena
respectively. We will use these latter expressions in
order to get some explicit expressions for some parameters
related to scalar density perturbations and relic gravitational waves.

\section{scalar and tensor perturbations}

\subsection{A general approach to quantum perturbations}

Inflation causes perturbations through the amplification of
quantum fluctuations, which are stretched to astrophysical scales
by the accelerated expansion. Inflation generates two types of
perturbations, the density
perturbations, which come from quantum fluctuations in the inflaton
field, together with the corresponding scalar metric
perturbation\cite{L80}, and relic gravitational waves which are
tensor metric fluctuations\cite{G75}. The former gives rise to
gravitational instability and acts as seed of structure
formation\cite{MFB92}, while the latter predicts a stochastic
background of relic gravitational waves.

The gauge invariant Mukhanov-Sasaki variable\cite{S86,M88,M05},
defined from $u = z \cal{R}$, with $z= a \frac{\dot{\phi}}{H}$ and
$\cal{R}$ corresponds to the gauge-invariant comovil curvature perturbation,
plays an prominent role in cosmology, not only because of its
simple relation with the perturbed scalar curvature on the
spatial hypersurfaces but also because, in general relativity,
as we will see, it behaves as a Klein–Gordon field. In terms of
the Fourier transformed of the Mukhanov-Sasaki variable the corresponding equation
results to be

\be \frac{d^2 u_k}{d \eta^2} + \left(k^2 -
\frac{1}{z}\frac{d^2z}{d\eta^2}\right) u_k=0, \label{27} \en
 where $\eta = \int{\frac{1}{a}\,dt}$ is the conformal time
 and $\frac{1}{z}\frac{d^2z}{d\eta^2}$ corresponds to the mass
 term.

 We have that during inflation $k^2 \gg
 \frac{1}{z}\frac{d^2z}{d\eta^2}$,
 and thus the latter equation can be solved to get at early time
 a solution of the type $ \displaystyle u_k(\eta) \sim e^{-ik\eta}\left(1+\frac{{\cal
 {A}}_k}{\eta}+....\right)$.

When $k^2  \ll
 \frac{1}{z}\frac{d^2z}{d\eta^2}$, it is found that the physical modes
 present wavelengths much bigger than the curvature
 scale.

In solving equation (\ref{27}), it is needed
to impose boundary conditions. Usually, the asymptotic
conditions are taken to be
\be
u_k \rightarrow \left\{ \begin{array}{lll}
\frac{1}{\sqrt{2k}} e^{-i k \eta} &\hspace{0.5cm} $as$& -k \eta
\longrightarrow \infty, \\
{\cal {A}}_k z &\hspace{0.5cm} $as$  & -k \eta \longrightarrow 0.
\end{array}\right.,
\label{30} \en
the so-called Bunch-Davies vacuum state\cite{K12}.
These conditions guarantee that perturbations that are generated well inside the
horizon, i.e. in the region where $k \ll aH$, the modes approach
plane waves, and those that are generated well outside the horizon,
i.e. in the region where $k \gg aH$, are fixed.

The primordial scalar perturbation is defined from the two point
correlation function, which results to be
\be {\cal{P}}_{\cal{R}}(k)=\frac{k^3}{2\pi^2}<
{\cal{R}}_{{\overrightarrow{k}'}} {\cal{R}}_{{\overrightarrow{k}}
}>
\delta (\overrightarrow{k}'+\overrightarrow{k} )=
\frac{k^3}{2\pi^2}\left|\frac{u_k}{z}\right|^{^{2}}.
\label{28}
\en

From the primordial scalar perturbations we can define
the scalar spectral index
\be n_s -1\equiv \frac{d \ln {\cal{P}}_{\cal{R}}}{d \ln {k}},
\label{ns}\en
which in terms of the hierarchy Hubble parameters and considering into
account higher order corrections, this reduces to\cite{EP06}
\be n_s - 1 = -4\epsilon_{_{H}} + 2 \eta_{_{H}}
-2(1+{\cal{C}})\epsilon_{_{H}}^2- \frac{1}{2}(3-5\,{\cal{C}})\epsilon_{_{H}}\eta_{_{H}}
+\frac{1}{2}(3-{\cal{C}})\xi_{_{H}}^2,\label{nss}\en
where ${\cal{C}}=4(\ln{2}+\gamma)-5$ with $\gamma=0.5772$ represents
the {\it Euler-Mascheroni constant}.

In the same order correction the {\it running scalar spectral index}, $\alpha_{s} \equiv
\displaystyle \frac{d n_s}{d \ln {k}} $ becomes
\be \alpha_{s}=\frac{1}{1-\epsilon_{_{H}}}\left[8\epsilon_{_{H}}^2
+ 10 \epsilon_{_{H}} \eta_{_{H}}- 2\,
\xi_{_{H}}^2-\frac{7\,{\cal{C}}-9}{2}\epsilon_{_{H}}\xi_{_{H}}^2+\frac{{\cal{C}}-3}{2}
\eta_{_{H}}\xi_{_{H}}^2\right].
\label{alphas} \en

On the other hand,
transverse-traceless tensor perturbations are generated
from quantum fluctuations during inflation\cite{MFB92}.
Tensor perturbations do not couple to matter and thus they
are  determined by the dynamics of the background metric only.
The tensor perturbations evolve
like minimally coupled massless fields whose spectrum becomes represented by
${\cal {P}}_{{\cal T}}$ and thus, we can introduce the {\it gravitational wave spectral index}
$n_{_{T}}$ given by $\displaystyle n_{_{T}} \equiv \frac{d \ln
{{\cal {P}}_{{\cal T}}}}{d \ln {k}}$, which becomes at second order correction\cite{SL93}
\be n_{_{T}} = -2\,\epsilon_{_{H}}-(3+{\cal{C}})\epsilon_{_{H}}^2+
(1+{\cal{C}})\epsilon_{_{H}}\eta_{_{H}}.
\label{nT}
\en
In the same way, we can introduce the {\it
running tensor spectral index},
$\alpha_{_T}\equiv \displaystyle \frac{d n_{_T}}{d \ln {k}}$ which results
into
\be
\alpha_{_T}=-4\frac{\epsilon_{_{H}}}{1-\epsilon_{_{H}}}\left(\epsilon_{_{H}}-\eta_{_{H}}\right)
-(1+{\cal{C}})\frac{\epsilon_{_{H}}\xi_{_{H}}}{1-\epsilon_{_{H}}}. \label{alphaT} \en

We define the
tensor-to-scalar amplitude ratio $\displaystyle r \equiv
\frac{{\cal {P}}_{{\cal T}}}{{\cal{P}}_{\cal{R}}}$ which results to be at second order correction
\be r = 16 \epsilon_{_{H}}\left[1+2\,{\cal{C}}\left(\epsilon_{_{H}}-\eta_{_{H}}\right)\right].
\label{r} \en

We want to determine these general expressions for the case of intermediate
inflation, which we describe in the next subsection.

\subsection{The intermediate inflationary case}

We would like to apply the previous approach to the intermediate
inflationary universe model, where our finality is to obtain the
different expression as a function the number of e-folds in order
to compare with the current observational data. In this context
this approach is similar to that described in Refs.\cite{M13} and
\cite{G-BR14}. Thus, in doing this, we use expressions
(\ref{epsilonN}) and (\ref{etaN}) in the different relevant
expressions for $n_s$, $\alpha_s$, $n_T$, $\alpha_T$ and $r$. Firstly, we
obtain for the parameter $n_s(N)$ the following expression
\be n_s(N) -1 = -2\left(\frac{1}{1-\frac{4
N}{\beta}}\right)\left[1-\frac{2}{\beta}+\left(1-(3+{\cal{C}})\frac{1}{\beta}
-2(3-{\cal{C}})\frac{1}{\beta^2}\right)\left(\frac{1}{1-\frac{4N}{\beta}}\right)\right].
\label{nsN}
\en
We can use this expression to obtain the corresponding value of
the parameter $f$ when the current data released by
planck\cite{P13}, which is $n_s=0.9603\pm 0.0073$, is considered
at some given e-folds number, $N$. For instance, if we take $N=60$
we get $f=0.03181 \pm 0.0005$, and if $N=70$ we obtain $f=0.02741
\pm 0.00006$. These values for $f$ correspond to $\beta \simeq 122
$ and $\beta \simeq 142$, respectively. However, we saw previously that
in order to have the
Hubble parameter real it is needed to satisfy that $\beta$ must be
greater than $4 N$ (see expression (\ref{HN})), and we observe
that these value for $\beta$
violate this condition, therefore, we can conclude that there not
exist value of $f$ which agrees with the value of $n_s$ released
by Planck together with an appropriated value of the number of e-folds.

From the previous expression (\ref{nsN}) we can obtain the running scalar
spectral index, which becomes
\be
\alpha_{s}(N)=-\frac{\left(\frac{\beta}{4 N}-1\right)}{\left(1-\frac{4
N}{\beta}\right)^2}\left[8\left(2+\frac{1}{\beta}-\frac{2}{\beta^2}\right)-\left(1+\frac{6}{\beta}
+\frac{8}{\beta^2}\right)\left(3-{\cal{C}}-2\beta(3-2{\cal{C}})\right)\left(\frac{1}{1-\frac{4 N}{\beta}}\right)\right].
\label{alphsN}
\en
Planck\cite{P13} satellite reported a value for this parameter
given by $\alpha_{s} = - 0.0134 \pm 0.0090$. Even though this
value is not statistically significant we can use it in order to
obtain some values of the parameter $f$, just the way we did with
$n_s$. Here, it is found that for $N=60$ and $N=70$ we get $f
\simeq 0.0133$ and $f \simeq 0.0114$, respectively. These values
for the parameter $f$ give rise to $\beta \simeq 297$ and $\beta
\simeq 347$, respectively. Although these values satisfy
the constraint $\beta > 4 N$ new report on the measurement of $\alpha_s$
could change this conclusion, since the value of the parameter $f$ is
very sensitive to the change of the value of the parameter $\alpha_s$.

On the other hand, the tensor spectral index becomes
\be n_{_{T}}(N) = -\left(\frac{2}{1-\frac{4
N}{\beta}}\right)\left[1+\left(1-\frac{1+{\cal{C}}}{\beta}\right)
\left(\frac{1}{1-\frac{4 N}{\beta}}\right)\right],\label{nTN}
\en

For a single-field inflationary model results a relation at first order
between
the parameters $r$ and  $n_{_{T}}$, the so called {\it consistency
relation}, which is expressed by
$n_{_{T}}=-\frac{r}{8}$\cite{LL92}. This relationship is due to
both scalar and tensor perturbations come from a single degree of
freedom, which is carried by the inflaton field. In the second order approximation
this consistency condition becomes more subtle\cite{FFM10}, but in order to get an estimate
is sufficient to consider the above relation. If we use the
value reported recently by BICEP2\cite{B14} for the $r$ parameter,
i.e. $r \simeq 0.2$, it is obtained ruffly
$n_{_{T}} \simeq -0.025$. With this value at hand we get from expression
(\ref{nTN}) that the values for the parameter $f$ at $60$ and $70$
e-folds become $f \simeq 0.0319$ (corresponding to $\beta \simeq
121$) and $f \simeq 0.0275$ (corresponding to $\beta \simeq 141$), respectively.
Similarly, what happens to the $n_s$ parameter, we find that
the inequality $\beta > 4 N$ is not met again.

With respect to the running tensor spectral index we find that it becomes in this case
\be \alpha_{_T}(N)=\left(\frac{2}{N}\right)\left[\frac{1}{8}(\beta
+ 2)(1+{\cal{C}})-1\right]\left(\frac{1}{1-\frac{4N}{\beta}}\right)
\label{alphTN} \en

Unfortunately, the running tensor index is poorly constrained with the current data set,
and thus, usually it is ignored and people constrain the different models using  as observable
$n_s$, $\alpha_{_{s}}$ and $r$ as free parameters. It is expected that upcoming experimental data
verify whether they actually are sizable, both tensor mode perturbations and the
running of the spectral index. Certainly, this will shed light on
the natural structure of the inflaton scalar potential.

\begin{figure}[th]
\centering
\includegraphics[width=12cm,angle=0,clip=true]{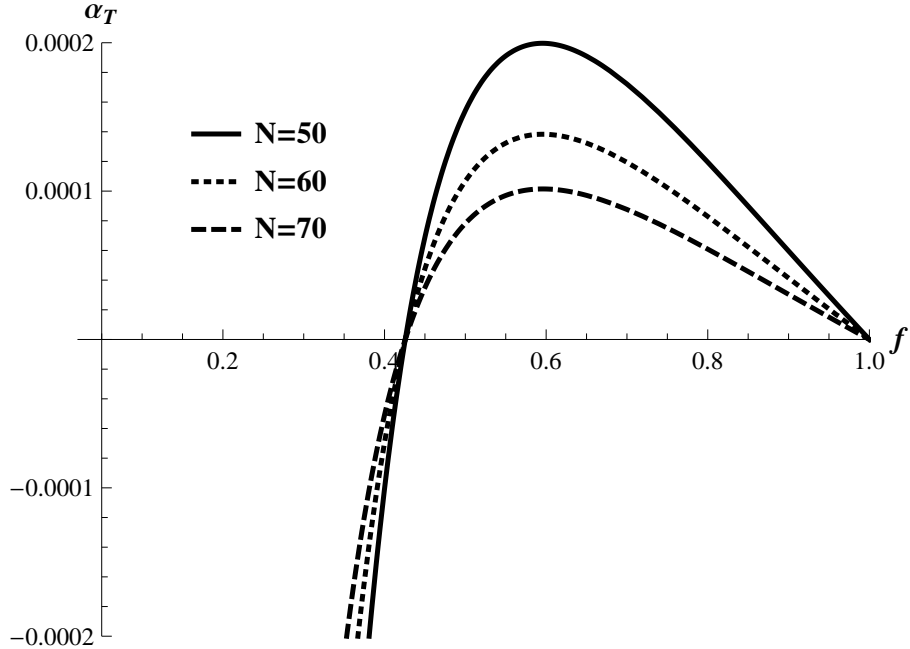}
\caption{This plot shows  the parameter $\alpha_{_{T}}$ as a function of the
free parameter $f$ for three values of the number of e-folds, $N=50$, $N=60$
and $N=70$.
} \label{fig1}
\end{figure}

Fig \ref{fig1} represents $\alpha_{_{s}}$ as a function of the parameter $f$. It
shows that the value $f \simeq 0.425$, or
equivalent $\beta \simeq 5.4$, is a value which divides what would
correspond to a redshift ($\beta < 5.4$), with $\alpha_{_{T}}$
negative, and a blueshift ($\beta > 5.4$), with $\alpha_{_{T}}$
positive. Since $\beta$ should be greater that $4 N$, we see that
this model favors a blueshift for the tensor spectral index.
However, following a procedure similar to that used in obtaining
the consistency condition, it is possible to fix the running
tensor spectral index at second order in terms of the $r$ and
$n_s$ parameters so that it results into
$\displaystyle \alpha_{_{T}}=\frac{r}{8}\left(\frac{r}{8} + n_s -
1\right)$\cite{FFM10}. From this expression we could obtain an estimation for
the running tensor spectral index if we use those values given by
Planck ($n_s \simeq 0.9603$) and BICEP2 ($r \simeq 0.2$). This
estimation results to be $\alpha_{_{T}} \simeq -0.0004$, which,
according to the graph \ref{fig1}, it corresponds to a redshift, and
thus, violating the inequality $\beta > 4 N $.

Let us now consider the tensor-to-scalar ratio $r$.
For intermediate inflation it becomes in terms of the number of e-folding
\be r(N) = \left(\frac{16}{1-\frac{4
N}{\beta}}\right)\left[1-\frac{4
{\cal{C}}}{\beta}\left(\frac{1}{1-\frac{4 N}{\beta}}\right)\right].
\label{rN} \en

\begin{figure}[th]
\centering
\includegraphics[width=12cm,angle=0,clip=true]{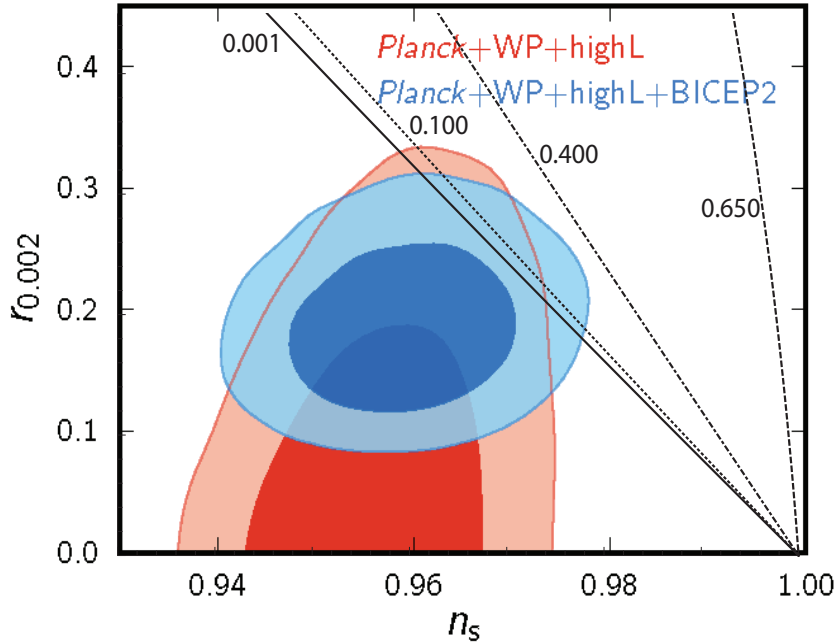}
\caption{This plot shows  the parameter $r$ as a function of the
scalar spectral index $n_s$ for four values of the constant $f$,
which correspond to $f =0.001; 0.100; 0.400; 0.650$, as described
by Eq. (\ref{rns}) by choosing the minus sign. These curves are
contrasted with the recent data released by Planck\cite{P13}
combined with the WMAP large-angle polarization and
BICEP2\cite{B14}. The $n_s/r$ plane showed here corresponds to
Fig.13 from Ref.\cite{B14}, in which the red contours are the
Monte Carlo Markov Chains provided with the Planck data release,
while the blue one correspond to the plots when the BICEP2 data
are taken into account. Each of both contours represent the $68\%$
and $95\%$ confidence regions. Note that this model is
disfavoured, being outside of the $95 \%$ contour region for any
value of the parameter $f$, or equivalently $\beta$. }
\label{fig2}
\end{figure}

Now, from expression (\ref{nsN})  we could get $N$ as a function of $1-n_s$ and then
replace this result into the previous expression, Eq. (\ref{rN}), and thus to obtain
the parameter $r$ as a function of $n_s$. This expression reduces to
\bea
r(n_s)^{\pm}=&\hspace{-3.5cm}  \left(\frac{8}{\delta}\right)\left(1-\frac{2}{\beta}\right)\left[1
\pm \sqrt{1
-\frac{2\delta}{\left(1-\frac{2}{\beta}\right)^2}(1-n_s)}
\right]\times \nonumber \\
&\hspace{2.5cm}
\left\{1-\frac{2 {\cal{C}}}{\beta \delta}\left(1-\frac{2}{\beta}\right)\left[1
\pm \sqrt{1
-\frac{2\delta}{\left(1-\frac{2}{\beta}\right)^2}(1-n_s)}\right]\right\},\label{rns}
\ena
where $\delta \equiv
1-\frac{3+{\cal{C}}}{\beta}-\frac{2(3-{\cal{C}})}{\beta^2}$. Here, with
the finality of obtaining $r=0$ when $n_s = 1$ we choose in the following the minus
sign. With this choice, it is not hard to see that for $\displaystyle
\frac{2\delta}{\left(1-\frac{2}{\beta}\right)^2}(1-n_s) \ll 1$ (which is equivalent to
consider just the first order approximation in the Hubble hierarchy parameters) it is obtained that\cite{BL93}
\be r(n_s) \thickapprox
\frac{8\,\beta}{(\beta-2)}\left(1-n_s\right). \label{rns01} \en
In Fig. \ref{fig2} we show the relation between $r$ and $n_s$
for four different values of the parameter $f$, according
to expression (\ref{rns}) (by taking the negative sign), whose curves
are correlated with existing measures made by BICEP2\cite{B14}, together with
the data released by Planck\cite{P13} combined with the WMAP large-angle polarization.
The $n_s/r$ plane showed here corresponds to Fig.13 of Ref.\cite{B14}, in
which the red contours are the Monte Carlo  Markov Chains provided
with the Planck data release, while the blue one correspond to the
graphs when the BICEP2 data are taken into account. Each of both contours represent
the $68\%$ and $95\%$ confidence regions for the corresponding parameters.
We notice from this graph that the curves which are in the range $0.7 \lesssim f < 1.0$
lie outside the contour of $68 \%$ confidence level.  But, with respect to the $95 \%$ confidence
level region this model is disfavoured, being outside of this
contour for any value of the parameter $f$, or equivalently $\beta$.

\section{conclusion}

We have studied single-field model of intermediate inflationary
universes. This study was made under the scheme of the
Hamilton-Jacobi approach, which mainly rests on the generating
function, which is given by the Hubble factor, $H$, as a function
of the scalar inflaton field, $\phi$. Within this scheme a series
of hierarchical parameters are defined in terms of the generating
function and its derivatives, the so-called hierarchy Hubble
parameters. Now, these hierarchy parameters satisfy the set of
flow equations, commonly called the hierarchy flow equations. This
set of equation corresponds to a set of lineal differential
equations, which are characterized by derivatives of the hierarchy
Hubble parameters with respect to the number of e-folds. By their
nature, all higher-level hierarchy Hubble parameters are
determined by the basic parameter, the first hierarchy Hubble
parameter and its derivatives, when it is expressed in terms of
the number of e-folds.

With the different hierarchy Hubble parameters at hand, we could get
at second order in these parameters the corresponding observational
cosmological parameters, such that the scalar spectral index, its
running and the tensor-to-scalar ratio, among others, related to intermediate
inflation. When these parameters are confronted with the current observational
data, such that Planck and BICEP2, it is found that this model is disfavoured,
since, for instance, in the $n_s/r$ plane all the curves are outside of the
$95 \%$ contour region for any value of the parameter $f$, or equivalently $\beta$.

Another drawback that this model presents is related to its scalar
potential. The main characteristic that the potential presents in
this model, it does not have a minimum (except for $\phi
\longrightarrow \infty$, where $V(\phi) \longrightarrow 0$, for any
value of $\beta$). This makes it impossible for inflation end
without the aid of an additional scalar field. In this respect, some authors
have introduced the so-called curvaton scalar field.

Before concluding we would like to mention that the intermediate
inflationary universe model is similar to other models in which
the scale factor lies between a power law and a de-Sitter phase,
such as the logamediate scenario, for example\cite{BN07,varios12}.
We guess that these sort of models might suffer of the same
problems than that found in the intermediate scenario when they
are study under the single-field approach. However, we believe
that this is a point that should be considered carefully.

\begin{acknowledgments}
This work was supported by the COMISION NACIONAL DE CIENCIAS Y
TECNOLOGIA through FONDECYT Grant N$^{0}$ 1110230 and also was
partially supported by PUCV Grant N$^0$ 123.710/2011.

\end{acknowledgments}

\end{document}